\documentclass[reprint,prb,showkeys,
superscriptaddress,citeautoscript,] {revtex4-1}

\usepackage{graphicx}
\usepackage{makecell}
\usepackage{multirow}
\usepackage{hyperref}
\hypersetup{colorlinks=true, urlcolor= blue, citecolor=blue, linkcolor= blue, 
bookmarks=true, bookmarksopen=false}

\usepackage{amsmath,amssymb}
\usepackage{textcomp}
\usepackage{multirow}

\usepackage{enumitem}
\setlist{noitemsep, leftmargin=*}

\usepackage{times}
\begin{document}
\title{Bilayer borophene: The effects of substrate and stacking}
\author{Shobair Mohammadi Mozvashi}

\author{Mojde Rezaee Kivi}

\author{Meysam Bagheri Tagani}
\email{m{\_}bagheri@guilan.ac.ir}
\affiliation{Department of Physics, University of Guilan, P. O. Box 41335-1914, 
	Rasht, Iran.}

\begin{abstract}

Bilayer borophene has recently attracted much interest due to its outstanding 
mechanical and electronic properties. The interlayer interactions of these 
bilayers are reported differently in theoretical and experimental studies. 
Herein, we design and investigate bilayer $\beta_{12}$ borophene, by 
first-principles calculations. Our results show that the interlayer distance of 
the relaxed AA-stacked bilayer is about 2.5 \AA, suggesting a van der Waals 
(vdW) 
interlayer interaction. However, this is not supported by previous experiments, 
therefore by constraining the interlayer distance, we propose a preferred model 
which is close to experimental records. This preferred model has one covalent 
interlayer bond in every unit cell (single-pillar). 
Further, we argue that the preferred model is nothing but the relaxed model 
under a 2\% compression. Additionally, we designed three 
substrate-supported 
bilayers on the Ag, Al, and Au substrates, which lead to double-pillar 
structures. Afterward, we investigate the AB stacking, which forms 
covalent bonds 
in the relaxed form, without the need for compression or substrate. Moreover, 
phonon dispersion shows that, unlike the AA stacking, the AB stacking is stable 
in freestanding form. Subsequently, we calculate the mechanical properties of 
the AA and AB stackings. The ultimate 
strengths of the AA and the AB stackings are 29.72 N/m at 12\% strain and 23.18 
N/m at 8\% strain, respectively. Moreover, the calculated Young’s moduli are 
419 N/m and 356 N/m for the AA and the AB stackings, respectively. These 
results show the superiority of bilayer borophene over bilayer MoS$_2$ in terms 
of stiffness and compliance. Our 
results can pave the way of future studies on bilayer borophene 
structures.

\end{abstract}

\maketitle

\section{Introduction}

Borophene has recently attracted a surge of interest for its outstanding 
electronic and mechanical properties \cite{ebrahimi22, li18, mozvashi21, 
mohebpour22, mogulkoc2021design}. It is the lightest 2D material, 
rendering it a promising candidate for lightweight nanodevices \cite{li15, 
saha2021fused}. 
Moreover, 
the electron deficiency of boron atoms causes complex bonding which in turn 
results in diverse allotropes for borophene, including $\alpha$, $\beta$, and 
$\gamma$ \cite{zhou16, chand21}. In addition to monolayer, bilayer 
borophenes have also attracted much 
attention. It was expected that bilayer borophene would be more stable than 
monolayer borophene due to the interlayer bonding \cite{xu22}.

To date, many theoretical and experimental works have been conducted on the 
subject of different bilayer borophene allotropes and their properties 
\cite{xu22, gao18, ma16, nakhaee18}. 
Moreover, there are still various questions to be answered. For instance, 
theoretical studies have suggested the interlayer distance of bilayer borophene 
in the range of 2.5 to 3 \AA, suggesting a van der Waals (vdW) interaction 
between 
the layers \cite{ahn20, zhong18, wu2021}. However, the synthesized 
bilayer borophenes show a much 
closer interlayer distance, around 2 \AA, implying relatively strong covalent 
bonds \cite{liu22, chen22}. However, some theoretical studies considered some 
constraints 
to design the bilayer borophenes with similar interlayer distance to the 
experiment \cite{gao18, li19}. Formation energies and phonon dispersions prove 
that the 
constrained models are more stable than the fully relaxed models.

In this paper, by first-principles calculations, we answer why the interlayer 
coupling in bilayer borophene should be covalent and under what conditions this 
occurs. We first investigate the bilayer $\beta_{12}$ borophene without 
constraining 
the interlayer distance, or \emph{``the relaxed model''}. Afterward, by 
applying 
the constraint of interlayer distance we reach a structure more similar to the 
experimental observations, phrased as \emph{``the preferred model''}. This 
model, 
which 
has one covalent interlayer bond in each unit cell, is more favorable than the 
relaxed model. Interestingly, by applying compressive strain on the relaxed 
model, it undergoes a transition to the preferred model and covalent bonds form 
between the layers. In other words, we suggest that the preferred model is 
nothing but the relaxed model under compression.

The experimentally stable bilayer borophenes were synthesized on a metal 
substrate with negative mismatches with borophenes, which apply a compressive 
strain on the overlayers \cite{liu22, chen22, sutter21}. Otherwise, the second 
boron layer does 
not grow regularly on the first one; instead small clusters of boron form 
\cite{chen22}. 
Thus, the fundamental factor for the stability of bilayer borophene and 
covalent 
interlayer bonds could be the compressive strain from the substrate. To prove 
this suggestion, we considered substrate-supported borophene bilayers on Al 
(111), Ag (111), and Au (111) surfaces and optimized the bilayer borophene on 
them. Our results show that the substrate-supported bilayer borophenes are more 
stable, with two interlayer covalent bonds in each unit cell. Our calculations 
show that these extra covalent bonds are due to the charge transfer from the 
substrate to the overlayers. This addresses well the question of why and how 
the bilayer borophenes can grow efficiently on the metal substrates and pave 
the way for future experiments. In other words, we suggest that the AA stacking 
of bilayer $\beta_{12}$ borophene requires a substrate with negative mismatch 
to be stable. On the other hand, the AB stackings of the 
bilayer $\beta_{12}$ borophene are stable and covalently bonded in the relaxed 
form. No 
compression nor substrate is needed. Therefore, we strongly suggest that the 
synthesis of an AB-stacked bilayer $\beta_{12}$ borophene is more probable than 
the AA-stacked one in freestanding form.

At last, we calculate and compare the mechanical properties of the AA and AB 
stackings. Our results show that the ultimate strength of the AA and the AB 
stackings are 29.72 N/m at 12\% strain and 23.18 N/m at 8\% strain, 
respectively. Moreover, Young’s moduli of the AA and the AB stackings are 419 
N/m and 356 N/m, respectively, which show higher stiffness and compliance of 
this bilayer 
compared to bilayer MoS$_2$. Generally speaking, in this paper, we tried to 
exploit the most 
needed mechanical and structural information about the AA and AB stackings of 
the bilayer $\beta_{12}$ borophene, to contribute in guiding the new 
explorations about this subject. 

\section{Computational details and equations}

The Spanish package solution, SIESTA \cite{siesta, siesta2}, was implemented 
for all the 
calculations, which is based on self-consistent density functional theory (DFT) 
and standard pseudopotentials. The exchange-correlation interactions were 
estimated through generalized gradient approximation (GGA), with 
parameterization of Perdew, Burke, and Ernzerhof (PBE) \cite{zhong18}. Based on 
the 
convergence of the total energy, as depicted in Supplementary Material Figs
S1 and S2, the reciprocal space was sampled by a mesh of 13×23×1~k~points in 
the Brillouin zone and the density mesh cut-off was set to 50 Ry. 
To consider the Van-der Waals interaction, the DFT-D3 correction of Grimme 
was implemented \cite{liu22}. Moreover, a vacuum space of 20 \AA\ was 
considered 
in the 
$z$-direction to prevent unwanted interactions. 

The interlayer binding energy of the freestanding bilayers was calculated 
through:

\begin{equation} \label{eq:Eb}
E_b=(E_{bi}-2E_{mono})/S
\end{equation}

\noindent where $E_{bi}$, $E_{mono}$, and $S$ are the total energy of the 
bilayer, total energy of each monolayer, and the area of the unit cell, 
respectively. The adhesion energy between the bilayer and the substrate was 
also calculated through:

\begin{equation}
E_{ad}=(E_{T}-E_{bi}-E_{sub})/S
\end{equation}

\noindent where $E_T$ is the total energy of the whole substrate-supported 
system and 
$E_{sub}$ is of the isolated substrate.

Young’s modulus is defined by:

\begin{equation} \label{eq:young}
Y_i=\frac{\partial\sigma_i}{\partial\epsilon_i}
\end{equation}

\noindent where $\sigma_i$ and $\epsilon_i$ are the stress and the strain in 
direction $i$. Also, $Y_{xy}$ is defined as the biaxial Young’s modulus. The 
stress tensor is explained in Supplementary Material Eq (S2). The $\sigma_{11}$ 
and $\sigma_{22}$ directly give the stress values for strains along the 
armchair and zigzag directions, respectively. For the biaxial strain, the mean 
square values of biaxial stresses were calculated by:

\begin{equation}
\sigma_{xy}=\sqrt{\frac{\sigma_{11}^2+\sigma_{22}^2}{2}}
\end{equation}

Moreover, the obtained stress values were multiplied by the vacuum distance (20 
\AA) to get the unit of N/m.

\section{Results and Discussion}

\subsection{The basic model}

\begin{figure*}
	\centering
	{\includegraphics[width=0.95\linewidth]{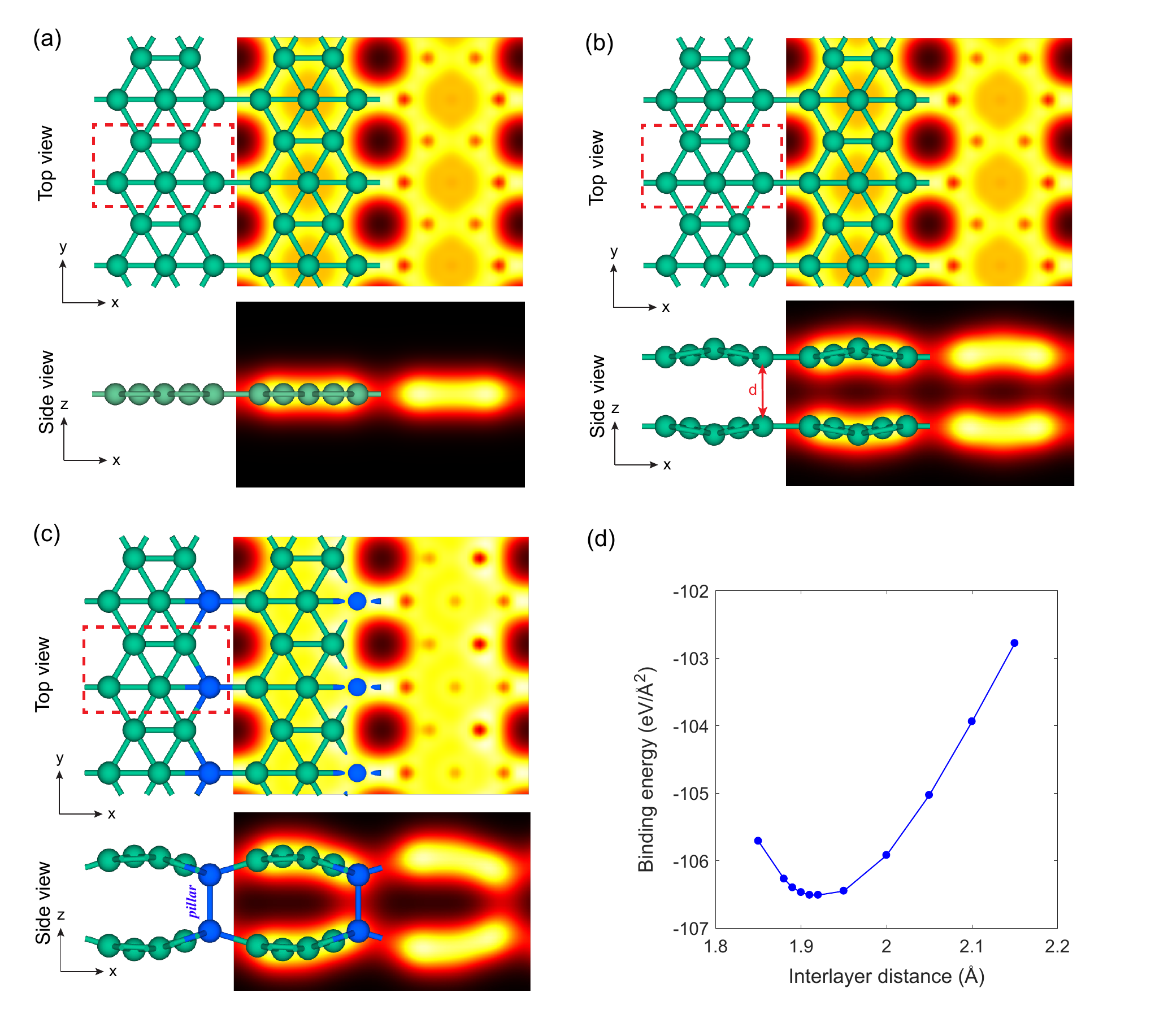}}
	\caption{Structural configuration and electron density map of \textbf{(a)} 
	monolayer $\beta_{12}$ borophene, \textbf{(b)} relaxed bilayer model, and 
	\textbf{(c)} preferred bilayer model. The unit cell, interlayer distance, 
	and pillar bonds are also depicted. \textbf{(d)} Binding energy as a 
	function of interlayer distance in the preferred model. The paces around 
	the minimum were smaller to obtain more precise answer.}
	\label{fig:Configuration}
    \end{figure*}

We start our investigations with monolayer $\beta_{12}$ borophene, which is 
shown in 
Fig \ref{fig:Configuration} (a). After full relaxation, a flat structure 
with 
lattice constants 
and average bond length of $a$ = 5.15, $b$ = 2.97, and $R$ = 1.74 \AA\ was 
obtained, 
which is consistent with previous theoretical and experimental records 
\cite{luo17, feng16}. Subsequently, we designed and optimized an AA-stacked 
bilayer, as shown in 
Fig \ref{fig:Configuration} (b). The lattice constants and the average bond 
length are $a$ = 5.14, 
$b$ = 2.98, and $R$ = 1.74 \AA. The binding energy using Eq (\ref{eq:Eb}) was 
calculated 
-99.5 
eV/\AA$^2$. The closest interlayer distance in this bilayer is $d$ = 2.45 \AA, 
which 
implies a van der Waals (vdW) interaction between the layers. We call this 
structure \emph{``the relaxed model''}, which is in agreement with several 
previous works 
\cite{ahn20, wu2021, zhong18}. However many stronger theoretical and 
experimental works suggest a 
closer interlayer distance ($\sim$ 2 \AA), and a covalent interlayer 
interaction 
for bilayer borophene \cite{liu22, chen22, li19}.

Here, two important questions arise: Why the interlayer interaction should be 
covalent? And why some of the theoretical works do not agree with the 
experiments? To address these questions, we considered a manipulated model, in 
which the interlayer distance was adjustable on demand. The procedure for 
designing and optimizing this model is described in Supplementary Material, 
Sec. S2. As described in Fig \ref{fig:Configuration} (c), we constrained this 
structure to have 
one interlayer covalent bond in each unit cell, labeled as \emph{`pillar'} and 
adjusted the interlayer distance to find the most stable state. Fig 
\ref{fig:Configuration} (d) 
shows the variation of binding energy as a function of interlayer distance in 
this configuration. Under these circumstances, the most favorable interlayer 
distance is around 1.91 \AA, with a binding energy of 106.5 eV/\AA$^2$. Also, 
the 
lattice constants and the average bond length are $a$ = 5.06, $b$ = 2.97, 
and R 
= 1.74 \AA, respectively. Interestingly, this configuration is more stable than 
the relaxed 
structure and it is more similar to the experiments, therefore, we call this 
structure \emph{``the preferred model''}.

In aspects of electronic properties, the relaxed and the preferred models share 
similar properties. As shown in Fig S7, they both are metals with dominant 
\textit{p} 
states around the Fermi level. However, as we saw in Fig 
\ref{fig:Configuration}, the electron 
density maps show different interlayer interactions for these two models. There 
are no electrons in the interlayer space of the relaxed model, which approves 
the weak vdW interaction between the layers. However, in the preferred model, 
we can see the presence of electron density between the so-called pillar 
atoms, which implies covalent-like interlayer bonds. The electron localization 
function (ELF) and the electron difference density maps are also available in 
Fig S8, approving this conclusion. All of these features suggest that the 
preferred model is more compatible with experimental studies \cite{liu22, 
chen22, li19}.

However, as mentioned above, the preferred model is unrealistic; for no one can 
hold the pillar atoms at a certain distance in real world. Then what makes this 
model so close to the experiment? The answer lies beneath the effects of the 
substrate. All the mentioned synthesized bilayer borophenes were grown on metal 
substrates, therefore, we should somehow take into account these effects. As we 
know, a substrate can influence the overlayers mechanically and electronically. 
The mismatch between the substrate and the overlayers can compress or stretch 
the latter, which affects other structural parameters including the interlayer 
distance. Moreover, a metal substrate, soaked in free electrons, can dope the 
overlayers to attract each other more strongly.

	\begin{figure*}
	\centering
	{\includegraphics[width=0.95\linewidth]{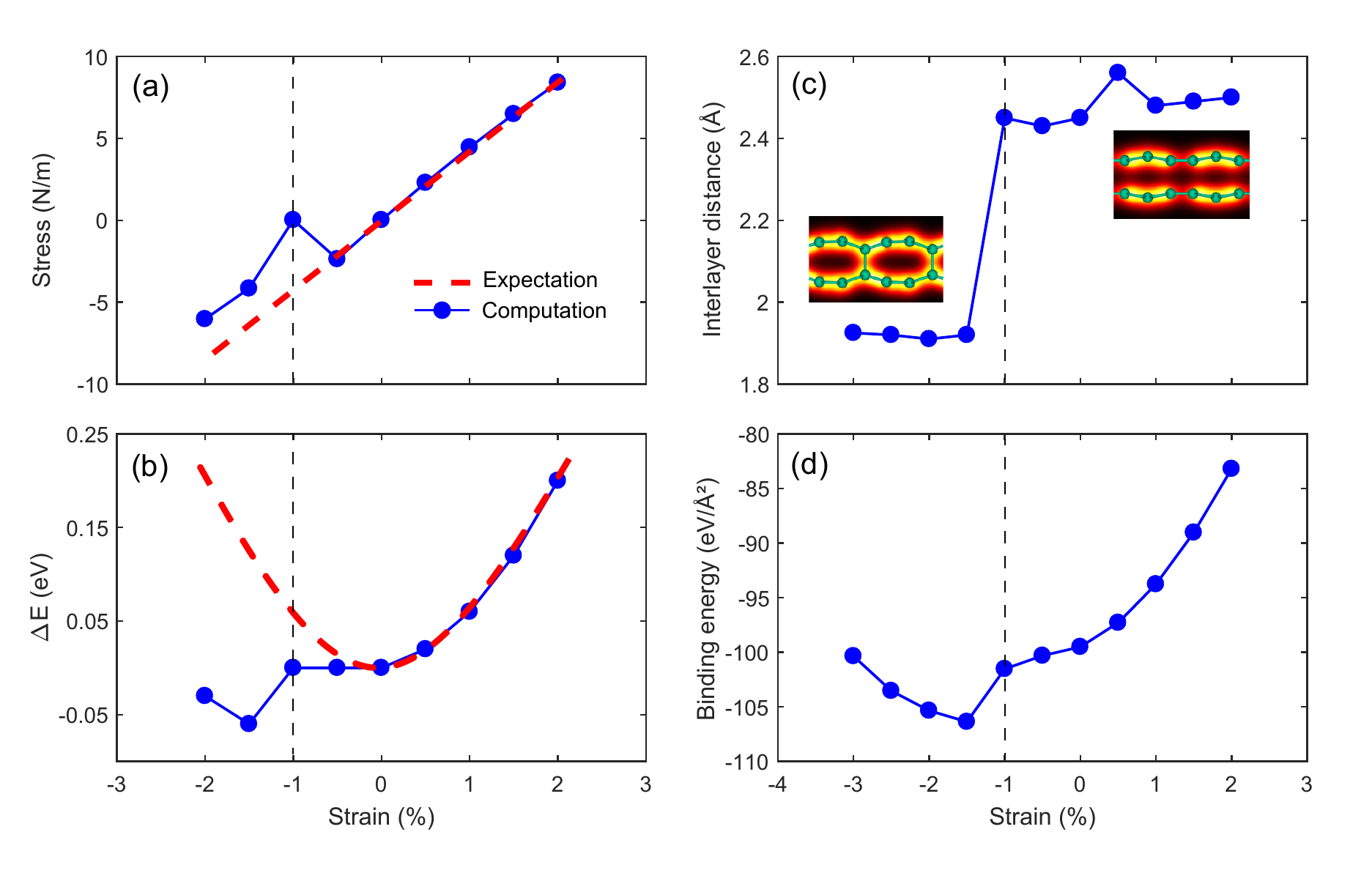}}
	\caption{Evaluation of \textbf{(a)} stress, \textbf{(b)} total energy, 
		\textbf{(c)} interlayer distance, and \textbf{(d)} binding energy of 
		the 
		relaxed model with the applied strain. The total energy was substituted 
		from the pristine total energy ($\Delta$E = E – E$_0$). The expected 
		harmonic behavior is shown with red dashed lines in (a-b).}
	\label{fig:Mech1}
\end{figure*}

We first simulate the mechanical effects of a possible substrate by applying 
biaxial strains on the relaxed model and observing the structural evaluation. 
We should keep an eye on the variation of stress and total energy with the 
applied strain to see if any structural phase transition takes place. In the 
harmonic range of a material, the stress is expected to behave linearly and the 
total energy to grow parabolic with compression or tension. The structural 
variations of the relaxed model with the applied strain are shown in Fig 
\ref{fig:Mech1}. 
After a 0.5\% compressive strain, the response of the stress and total energy 
deviates from the expected harmonic behavior, which implies a structural phase 
transition. Moreover, when the relaxed model is compressed around 1.5\%, the 
binding energy and the interlayer distance drop to -106.3 eV/\AA$^2$ and 1.93 
\AA, 
respectively, which is precisely consistent with the preferred model. In other 
words, the relaxed model turns into the preferred model under more than 1.5\% 
compression. This phase transition can be seen graphically in the insets of 
Fig \ref{fig:Mech1} 
(c). This explains well the successful synthesis of bilayer $\beta_{12}$ 
sheet on Ag 
(111), Al (111), and Au (111) substrates, all of which have mismatches between 
-1\% to -3\% with borophene \cite{liu22, chen22}. As we will further show 
by phonon 
dispersion, the AA-stacked bilayer $\beta_{12}$ is not stable in freestanding 
form, therefore, consideration of a proper substrate is inevitable.

\subsection{Substrate effects}

To explore the electronic effects of the substrate, we designed the 
substrate-supported bilayer models on the Ag (111), Al (111), and Au (111) 
surfaces. We used five-layers of substrate from which the top two layers were 
allowed 
to relax and the bottom three were fixed. Besides, the borophene overlayers 
were allowed to fully relax. 
These substrates, which are among the most frequently used surfaces for 
borophene 
synthesis \cite{zhong2017synthesis, yu2021viable}, all apply compressive 
strains into the borophenes due to their 
negative mismatch of the lattice constant. The lattice constants of the 
optimized substrate-supported models are $a$ = 5.00, $b$ = 2.88 \AA\ (Ag), $a$ 
= 
4.95, $b$ = 2.86 \AA\ (Al), and $a$ = 4.99, $b$ = 2.88 \AA\ (Au), which apply 
compressive 
strains of 3\%, 4\%, and 3\% to the bilayer, respectively. The interlayer 
distance in Ag-, Al-, and Au-supported models drops to 1.90, 1.84, and 2.2 \AA, 
which causes two covalent interlayer bonds in each unit cell to form. In other 
words, the compressive strain and the electrons transferred from the substrates 
cause the substrate-supported models to decrease the interlayer distance and 
make double-pillar covalently bonded bilayers as shown Fig S10. The electron 
doping facilities the interlayer bonding, therefore the substrate-supported 
models have one pillar more than our freestanding preferred model.

\begin{figure*}
	\centering
	\includegraphics[width=0.95\linewidth]{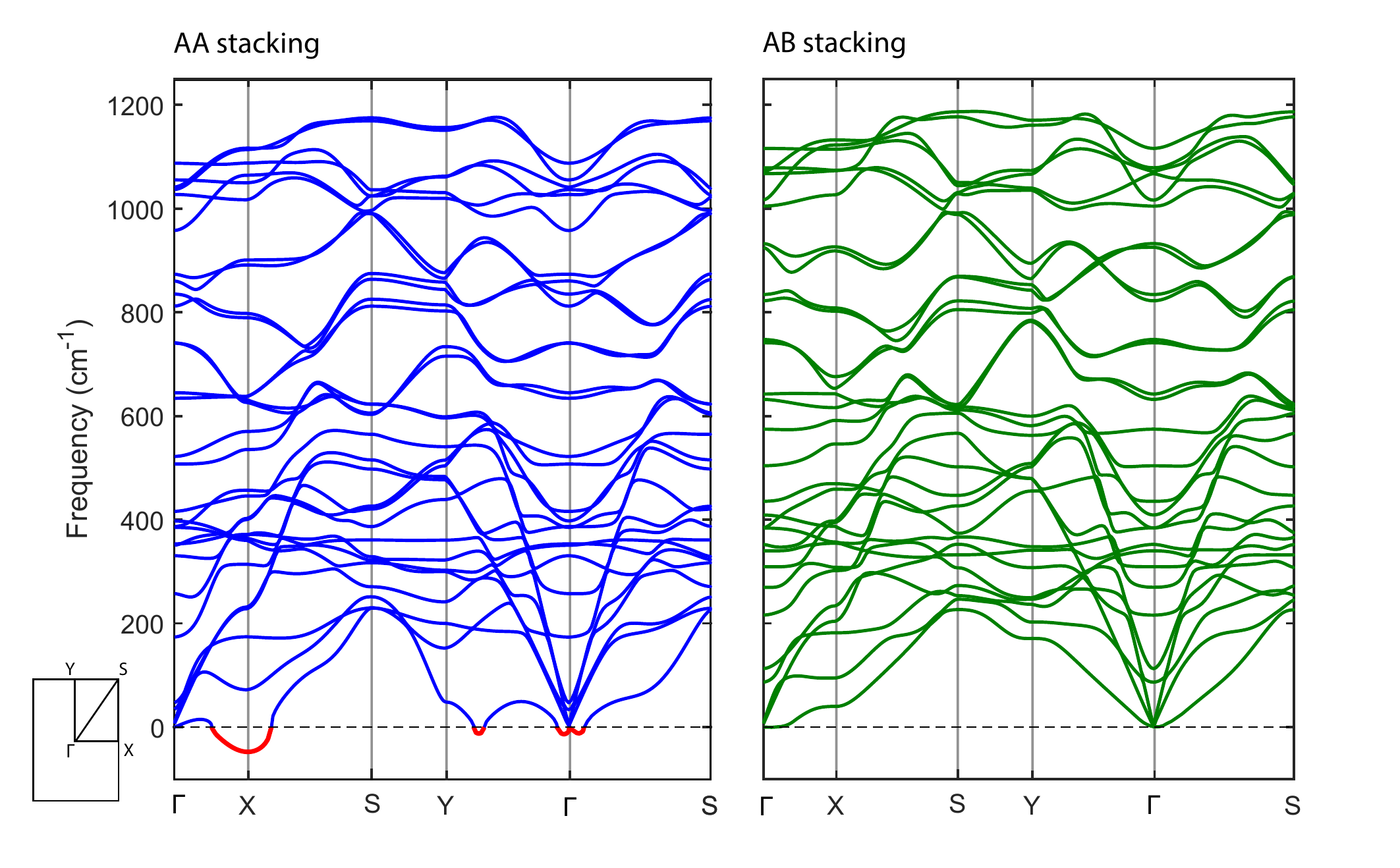}
	\caption{Phonon dispersion of bilayer $\beta_{12}$ borophene with AA (left) 
		and AB (right) stackings. The Brillouin zone is also depicted in the 
		inset.}
	\label{fig:phonon}
\end{figure*}

For a better understanding of this electron transfer, we calculated of Mulliken 
population of electrons among the layers. In all the substrate-supported 
models, the substrates donate and the overlayers accept electrons. In the 
Ag-supported model, the substrate averagely donates 7$\times10^{17}$ e/m$^2$ 
to 
the 
overlayer, where the lower and the upper layers averagely accept their shares 
as 3$\times10^{17}$ and 4$\times10^{17}$ e/m$^2$, respectively. These 
results support the 
previous 
theoretical study of borophene bilayer on Ag (111) substrate \cite{xu22}. In 
the 
Al-supported model the substrate donates an average of 3$\times10^{17}$ 
~e/m$^2$, from which 
the lower and the upper layers accept 1$\times10^{17}$ and 
2$\times10^{17}$~e/m$^2$, respectively. 
Moreover, in the Au-supported model the substrate donates around 12 
$\times10^{17}$ e/m$^2$ 
from which the upper and the lower borophenes accept 5$\times10^{17}$ and 
7$\times10^{17}$ e/m$^2$, 
respectively.

The injection of electron from the substrate compensates the 
electron deficiency of the borophene, providing a good condition for the two 
layers to make covalent bonds. The upper layer only makes bonds with the lower 
layer whereas the lower layer pays more electrons to make bonds with both the 
substrate and the upper layer, therefore the Mulliken population of the upper 
layer is higher. This also suggests for likelihood of production of more than 
two-layer structures, opening a way to obtain the bulk layered boron. We also 
calculated the adhesion energy between the substrates and the overlayer, which 
are 0.18, 0.15, and 0.22 eV/\AA$^2$ for Ag-, Al-, and Au-supported models, 
respectively. These values 
show more 
adhesion in comparison with the 
$\nu_{1/12}$ borophene on the Ag (111) substrate (0.11 eV/\AA$^2$) \cite{xu22}. 
For better 
comparison of 
the proposed substrate-supported models please see Table \ref{table:config}.

\begin{table}
	\centering
		\caption{Structural properties of the substrate-supported bilayer 
		$\beta_{12}$ borophenes: lattice constants ($a$ and $b$), interlayer 
		distance ($d$), substrate adhesion energy ($E_{ad}$), number of 
		covalent 
		bonds per unit cell ($n_B$), and density of electrons donated from the 
		substrate ($\rho_d$).}
	\
	\\
	\label{table:config}
	\begin{tabular}{lcccccc}
			\hline \hline
		Substrate & $a$ (\AA) & $b$ (\AA) & d (\AA) & $E_{ad}$ (eV/\AA$^2$) & 
		$n_B$ 
		& 
		$\rho_d$ ($10^{17}$ e/m$^2$)  \\
		\hline
		Ag (111)  & 5.00  & 2.88  & 1.90  & 0.18                          & 2  
		& 7                                 \\
		Al (111)  & 4.95  & 2.86  & 1.84  & 0.15                          & 2  
		& 3                                 \\
		Au (111)  & 4.99  & 2.88  & 2.20  & 0.22                          & 2  
		& 12    \\
		\hline \hline                            
	\end{tabular}
\end{table}

\subsection{Stacking effects}

Up to this point, we only concentrated on the AA stacking of $\beta_{12}$ 
bilayer 
borophene as the basic configuration. To take into account the effects of layer 
rotation, we turn our attention to the AB stacking, where the 6-folded B atoms 
of the upper layer are placed above the hexagon holes of the lower one, as 
shown in Fig S11. After full relaxation, this structure has the lattice 
constants of $a$ = 5.03 \AA, $b$ = 2.97 \AA, the average bond length of $R$ = 
1.76 
\AA, and 
binding energy of -114.43 eV/\AA$^2$. Interestingly, unlike the AA stacking, 
the 
AB stacking has a covalent interlayer interaction in the relaxed form with $d$ 
= 
2 \AA. Thus, the AB stacking does not need the support of a substrate to have 
covalent interlayer bonds.
The presence of covalent bonds among the layers is expected to improve the 
stability of the bilayer. To see these effects, we calculated phonon 
dispersions of both relaxed AA and AB stackings and compared them in 
\ref{fig:phonon}. 
The AA stacking has several negative modes with values of tens of cm$^{-1}$, 
which 
are signatures of dynamical instability. Interestingly, in the AB stacking 
phonon bands, no imaginary modes are seen, which approves its high stability. 
This suggests that, in a potential experiment, the bilayer $\beta_{12}$ 
borophene is 
very likely to grow with AB stacking in freestanding form. For a better 
comparison between the AA and AB stackings, please pay attention to Table 
\ref{table:models}.

Despite the differences in the interlayer bonding, the AA and AB stackings 
share most of the electronic properties including metallicity and orbital 
composition in density of states near the Fermi level (Fig S12). For the 
importance of the 
mechanical properties for applications of 2D materials, in the following 
section, we report and compare the mechanical properties of the bilayer 
$\beta_{12}$ borophene with the AA and AB stackings.

\begin{figure*}
	\centering
	\includegraphics[width=0.95\linewidth]{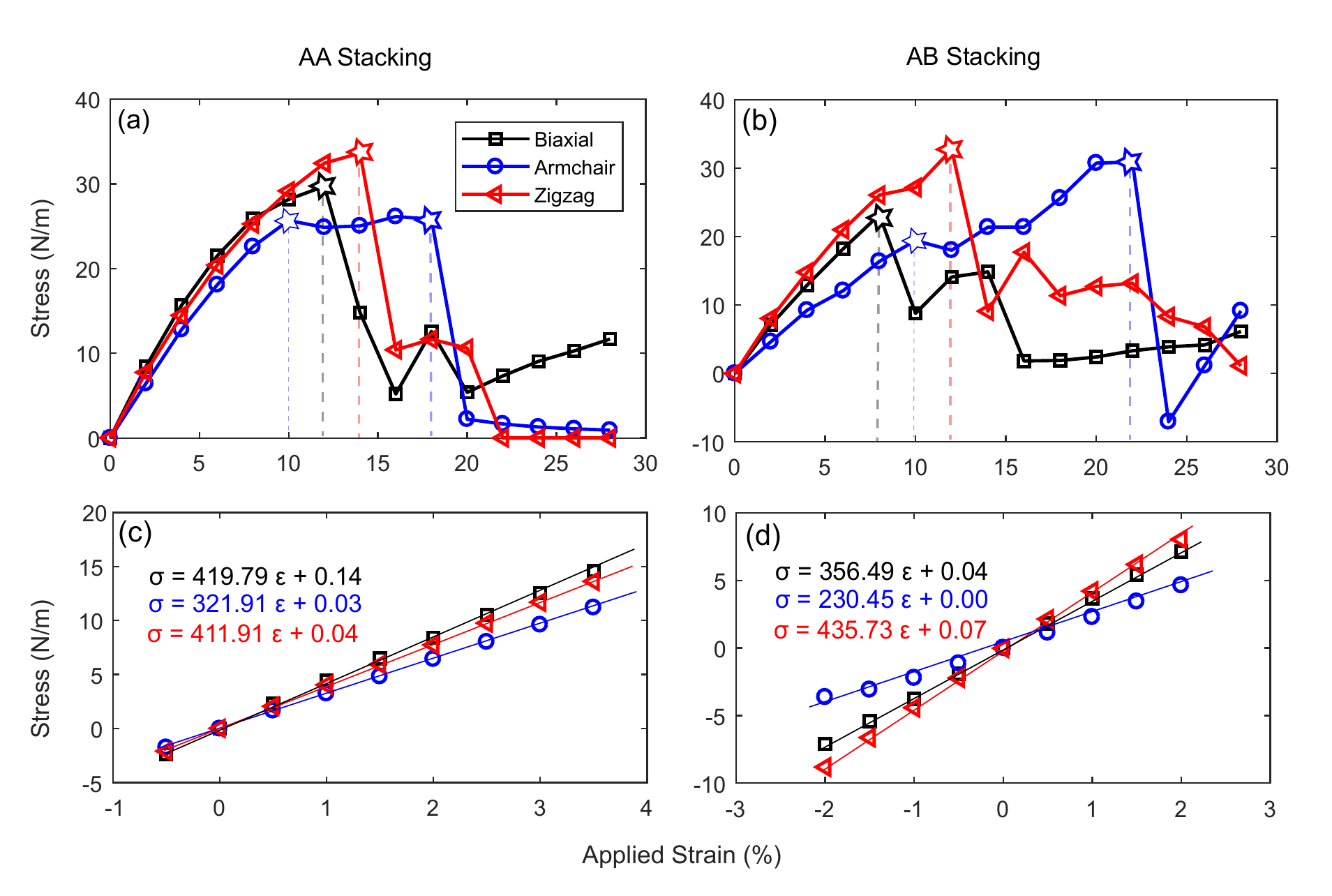}
	\caption{Mechanical properties of $\beta_{12}$ borophene with AA and AB 
		stacking. (a, b) Long-range stress-strain curves used to find the 
		critical 
		strains and ultimate strengths. (c, d) Short-range stress-strain curves 
		in 
		the harmonic range used to calculate Young’s moduli.}
	\label{fig:mech2}
\end{figure*}

\begin{table}
	\centering
	\caption{Comparison between different models in the freestanding form: 
	lattice constants ($a$ and $b$), interlayer distance ($d$), binding energy 
	($E_b$) and number of interlayer bonds in a unit cell ($n_b$).} \
\label{table:models}
	\\
	\begin{tabular}{lllllll}
		\hline \hline
		\multicolumn{2}{c}{Model}                 & $a$ (\AA) & $b$ (\AA) & $d$ 
		(\AA) 
		& $E_b$ 
		(eV/\AA$^2$) & $n_b$  \\ \hline
		\multirow{2}{*}{\rotatebox{0}{AA}} & Relaxed  & 5.14  & 2.98  & 
		2.45  
		& 
		99.53                        & 0   \\
		& Preferred & 5.06  & 2.97  & 1.91  & 106.52                       & 
		1   \\
		\multicolumn{2}{c}{AB}                    & 5.03  & 2.97  & 2.00  & 
		114.43                       & 1  \\ \hline \hline
	\end{tabular}
\end{table}

\subsection{Mechanical properties}

Knowing the mechanical properties of a material is very vital for its 
applications in nanodevices. Here, we report critical strains, ultimate 
strengths and Young’s moduli for bilayer $\beta_{12}$ borophene with AA and AB 
stackings. Here, the relaxed structures were used and no constraints were 
applied on the interlayer distances. We first applied tensile strain in the 
range of 0 to 30\% to evaluate the mechanical strength of the structures. As 
shown in 
Fig \ref{fig:mech2} (a), with biaxial strain in the range of 0 to 12\%, the 
stress of AA 
stacking rises to 29.72 N/m. Afterward, it suddenly drops to lower values. This 
gives us the critical strain and the ultimate strength under biaxial tension. 
For the zigzag direction, a critical strain of 14\% gives the ultimate strength 
of 33.63 N/m, mildly higher than the biaxial ones. In the case of the armchair 
direction, we have two critical strains of 10\% and 18\%, which give closely 
equal yield and ultimate strengths \cite{walker2014, goli2021dft} of around 26 
N/m.

A similar investigation was done for the AB stacking, shown in Fig 
\ref{fig:mech2} (b). 
The critical biaxial strain of 8\% gives an ultimate strength of 23.18 N/m. 
Moreover, the zigzag direction comes with critical strain and ultimate strength 
of 12\% and 32.74 N/m, respectively. Again, the armchair direction has two 
critical strains of 10\% and 22\%, with yield and ultimate strengths of 19.41 
N/m and 30.89 N/m, respectively. Regardless of the stacking, we can see that 
$\beta_{12}$ bilayer has a more complicated mechanical behavior in the armchair 
direction. This might be due to the more complex bonding characteristics in 
this direction.

Defined by Eq (\ref {eq:young}), Young’s modulus is the 
gradient of the 
stress-strain 
relation in the harmonic range. The harmonic range for the AA and AB stacking 
is 0\% to 4\% and -2\% to +2\%, respectively. The unconventional harmonic range 
of the AA stacking is due to the structural phase transition at -0.5\%, which 
was discussed before. As shown in Fig \ref{fig:mech2} (c, d), Young’s moduli 
of 
the AA 
stacking are 420, 322, and 412 N/m for biaxial, armchair, and zigzag strains. 
In a similar order, the AB stacking comes with 356, 230, and 436 N/m, 
respectively. Thus, the AB stacking is softer along the biaxial and armchair 
directions, but mildly stiffer along the zigzag direction. We compared the 
obtained mechanical properties with graphene, BN, and MoS$_2$ bilayers in Table 
\ref{table:mech}. Overall, we suggest that the stiffness and 
compliance of 
$\beta_{12}$ bilayer borophene is higher than 
MoS$_2$, but lower than graphene and BN.  
\begin{table}
	\centering
	\caption{Mechanical properties of AA and AB-stacked bilayer 
	borophene, compared with other 2D bilayers including graphene, BN, and 
	MoS$_2$: Young’s moduli ($Y_{ii}$), critical strain ($\epsilon$*), and 
	ultimate strength ($\sigma$*).}
\label{table:mech}
	\begin{tabular}{cccccc}
		\hline \hline
		 & Yx (N/m) & Yy (N/m) & Yxy (N/m) & $\epsilon$* (\%) & 
		$\sigma$* (N/m)  \\ \hline
		$\beta_{12}$ bilayer (AA)       & 322      & 412      & 420       & 
		12      & 29.7     \\
		$\beta_{12}$ bilayer (AB)       & 356      & 230      & 436       & 
		8       & 23.2    \\ 
		Bilayer graphene \cite{falin2017}  &--&--&620&11&72.1\\
		Bilayer BN \cite{falin2017}  &--&--&560&12&42.4\\
		Bilayer MoS$_2$ \cite{bertolazzi2011} &--&--&260&10&28.0\\
		\hline \hline
	\end{tabular}
\end{table}

\section{Conclusion}

In summary, by first-principles calculations, we investigated the bilayer 
$\beta_{12}$ borophenes with different structures. We suggest that the 
interlayer bonding plays an important role in the stability of the bilayer.
The AA stacking cannot make covalent interlayer bonds spontaneously, therefore 
it cannot grow in freestanding form. It requires a metal substrate such as Ag 
(111), Al (111), and Au (111) to be stable. These substrates, by 
applying compressive strain and doping electrons, help the two boron 
layers to attract each other more closely and make interlayer bonds. However, 
the AB 
stacking has covalent interlayer bonds which makes it stable in freestanding 
form. This is approved by phonon dispersion analysis. We also calculated the 
mechanical properties of the AA and AB stackings, which show higher stiffness 
and compliance of bilayer $\beta_{12}$ borophene than bilayer MoS$_2$. This 
results can 
give a positive contribution for future explorations about bilayer borophene 
structures.

\section*{Acknowledgment}

We are thankful to the Research Council of the University of Guilan for the
partial support of this research.

\section*{Declaration of Interests}
The authors declare that they have no conflict of interest.

\bibliography{Ref}
\end{document}